\documentclass[12pt]{article}
\usepackage{amsmath,amssymb}    

\usepackage{amsfonts}
\usepackage{enumerate}
\usepackage{ytableau}
\usepackage{hyperref}
\usepackage{bm}
\usepackage{tikz}
\usepackage{blkarray}  
\begin{document}
	\def\be{\begin{equation}}
	\def\bea{\begin{eqnarray}}
	\def\ee{\end{equation}}
	\def\eea{\end{eqnarray}}
\def\d{\partial}
\def\eps{\varepsilon}
\def\la{\lambda}
\def\b{\bigskip}
\def\nn{\nonumber \\}
\def\p{\partial}
\def\t{\tilde}
\def\h{{1\over 2}}
\def\be{\begin{equation}}
\def\bea{\begin{eqnarray}}
\def\ee{\end{equation}}
\def\eea{\end{eqnarray}}
\def\b{\bigskip}
\def\u{\uparrow}

\def\({\left(}
\def\){\right)}
\def\[{\left[}
\def\]{\right]}
\def\llangle{\langle\langle}
\def\rrangle{\rangle\rangle}
\def\ta{\theta}
\def\p{\partial}
\def\sgn{\mbox{sgn}}
\def\s#1{|_{#1}}
\def \ttb {$T\bar{T}$ }
\def \zb {\bar{z}}
\def \wb {\bar{w}}
\newcommand{\tabincell}[2]{\begin{tabular}{@{}#1@{}}#2\end{tabular}}
\def\be#1\ee{\begin{equation}\begin{split}#1\end{split}\end{equation}}
\def\({\left(}
\def\){\right)}
\def\[{\left[}
\def\]{\right]}
\def\lb{\left\{}
\def\rb{\right\}}
\def\f{\frac}
\def\la{\lambda}
\def\om{\Omega_{T}}
\def\omp{\Omega_{P}}
\def\ep{\epsilon}
\def\si{\sigma}
\def\p{\partial}
\def\pb{\bar{\partial}}
\def\pp{\partial\vec{\phi}}
\def\pbp{\bar{\partial}\vec{\phi}}
\def\vp{\vec{\phi}}
\def\vH{\vec{H}}
\def\ttb{T\bar{T}}
\def\va{\vec{\alpha}}
\def\cd{\cdot}
\def\s#1{|_{#1}}
\def\n#1{\vec{\nabla}{#1}}
\def\Tr{\text{Tr}}
\def \ttb {$T\bar{T}$}
\def \qb {\bar{q}}
\def \qA {\bar{A}}
\def \hs{\hspace}

\title{\textbf{Lax Connections in \ttb-deformed Integrable Field Theories}}
\vspace{14mm}
\author{Bin Chen$^{1,2,3}$, Jue Hou$^1$ and Jia Tian$^4$
\footnote{bchen01@pku.edu.cn, houjue@pku.edu.cn, wukongjiaozi@ucas.ac.cn}
}
\date{}
\maketitle

\begin{center}
	{\it
		$^{1}$School of Physics, Peking University, No.5 Yiheyuan Rd, Beijing 100871, P.~R.~China\\
		\vspace{2mm}
		$^{2}$Collaborative Innovation Center of Quantum Matter, No.5 Yiheyuan Rd, Beijing 100871, P.~R.~China\\
		$^{3}$Center for High Energy Physics, Peking University, No.5 Yiheyuan Rd, Beijing 100871, P.~R.~China\\
		$^4$Kavli Institute for Theoretical Sciences (KITS),\\
		University of Chinese Academy of Science, 100190 Beijing, P.~R.~China
	}
\vspace{10mm}
\end{center}

\makeatletter
\def\blfootnote{\xdef\@thefnmark{}\@footnotetext}  
\makeatother

\begin{abstract}
In this work, we try to construct the Lax connections of \ttb-deformed integrable field theories in two different ways. With reasonable assumptions, we make ansatz and find the Lax pairs  in the \ttb-deformed affine Toda theories and the principal chiral model by solving the Lax equations directly. This way is straightforward but maybe hard to apply for general models. We then make use of the dynamical coordinate transformation to read the Lax connection in the deformed theory from the undeformed one. We find that once the inverse of the transformation is available, the Lax connection can be read easily. We show the construction explicitly for a few classes of scalar models, and find consistency with the ones in  the first way. 
\end{abstract}

\baselineskip 18pt
\newpage

\tableofcontents

\section{Introduction}
The \ttb-deformation of two-dimensional field theories \cite{ttb1,ttb2}  has recently attracted much attention. It is a kind of solvable irrelevant deformation, and induces a flow in the space of field theories that satisfies the differential equation
\bea 
\p_t \mathcal{L}^{(t)}=\text{det}\({T_{\mu\nu}^{(t)}}\),
\eea
where $T_{\mu\nu}^{(t)}$ is the stress-energy tensor and $T\bar{T}=-\pi^2 \text{det}{T_{\mu\nu}^{(t)}}$. A remarkable property of this flow is that it preserves integrability if the undeformed theory is integrable. In the original paper \cite{ttb1}, the preservation of integrability under a \ttb-deformation or its generations has been supported by showing that the infinite conserved charges of the undeformed theory are still conserved under the flow. Another piece of evidence for integrability is from the fact that the S-matrix in the deformed theory is only modified by adding a CDD-like factor \cite{ttb1,cdd}.  A word of caution is that the solvability of the deformed theories does not rely on integrability crucially and it can be understood from various aspects \cite{aspect1,aspect2,aspect3,aspect4,aspect5,CT1}. Nevertheless, integrability may provide additional convenient handles on the theory. 

As it is well-known, integrability can be described in other frameworks, such as the Lax pair formulation and the B\"{a}cklund transformation formulation. In particular, the existence of the Lax connection is usually taken as the hallmark of classical integrability, and it also paves the way to quantization \cite{Sklyanin}.
However the Lax connection is notoriously difficult to find. Most of the time it needs the art of guess and trial and error. In this work, we will derive the Lax connections of several \ttb-deformed integrable theories with two different methods. The first method is kind of straightforward. We start from some reasonable ansatz, and find the connection by imposing equations of motion and solving the Lax equation. This method is suggestive but could be limited to specific models. The second method is more systematic, and it relies on the fact that the \ttb-deformation could be realized as a dynamical coordinate transformation\cite{CT1}. It is reminiscent of the method in deriving the Lax connections of $\gamma$-deformed superstring theory \cite{gamma}. This similarity is also expected, considering the fact that the holographic \ttb-deformation \cite{singletrace}, or known as the single trace \ttb-deformation, as well as the $\gamma$ deformations, can be related to a TsT deformation \cite{TsT1,TsT2}. The difference is that the single trace \ttb-deformation is a field redefinition while the \ttb-deformation of field theories is a change of coordinates.

The paper is organized as follows: In Sect. 2 and Sect. 3, we derive the Lax connection of (affine) Toda field theories and principle chiral model directly with reasonable ansatz; in Sect. 4 we first review the dynamical coordinate transformation approach of \ttb-deformation, and then reproduce the results obtained in section 2 and 3, and in the end we derive the Lax connection for a \ttb-deformed non-relativistic non-linear Schr\"{o}dinger theory; in Sect. 5 we end up with conclusions.

\section{\ttb-deformed (affine) Toda field theories}

In this section we consider the (affine) Toda field theories  and their \ttb-deformations. For the (affine) Toda field theories, the integrability can be studied from the point of view of  the Lax connections. As examples of $N$ scalar theories, the \ttb-deformed Lagrangians of these models have been derived in \cite{closed}. Here we are going to derive the deformed Lax connections with some reasonable ansatz.

\subsection{Undeformed theories}

The Lagrangian of a rank-$r$ affine Toda field theory\footnote{For a review on Toda field theory,  see \cite{Toda}.} is given by 
\be
\mathcal{L}^{(0)}&\equiv \pp\cd\pbp+V
\ee
with 
\be
V=-\f{m^2}{\beta^2}\sum_{i=0}^r n_i e^{\beta \va_i\cd\vp},
\ee
where $\vp$ is a vector field of $r$ components, the set of integer number $\{n_i\}$ characterizes the theory, $\{\va_i,~i=1,\dots,r\}$ are positive simple roots of the underlying Lie algebra and $\va_0=-\sum_i^r n_i\va_i$.
If in the summation the term $i=0$ is omitted then the theory reduces to the conformal Tada field theory.
The generators of the Cartan subalgebra $\vH=\{H_a,\ a=1,2,...,r\}$, and the simple roots $\{E_{\va_i},E_{-\va_i},\ i=0,1,...r\}$ satisfy the standard commutation relations 
\be
&\[H_a, H_b\]=0,\quad \[\vec{H}, E_{\pm\va_i}\]=\pm\va_iE_{\pm\va_i},\quad \[E_{\va_i}, E_{-\va_j}\]=\delta_{ij} \frac{2\va_j\cd\vec{H}}{|\va_j|^2},\\
&\[E_{\va_i}, E_{\va_j}\]=
\left\{
\begin{aligned}
\mathcal{N}_{\va_{i}+\va_{j}}E_{\va_{i}+\va_{j}},\text{\ \ \ \ if $\va_i+\va_j$ is a root},\\
0,\quad \text{\quad\quad\quad\quad if $\va_i+\va_j$ is not a root}.\\
\end{aligned}
\right.
\ee
The equations of motion are simply given by 
\be
2\p\pbp-\frac{\delta V}{\delta\vp}=0, 
\ee
and the Lax connections are
\be\label{Lax1}
L&=-\frac{\beta}{2}\pp\cd\vH-\la\sum_{i=0}^r m_i  e^{\beta \va_i\cd\vp/2}E_{\va_i},\\
\bar{L}&=\frac{\beta}{2}\pbp\cd\vH+\f{1}{\la}\sum_{i=0}^r m_i  e^{\beta \va_i\cd\vp/2}E_{-\va_i},
\ee
where $\la\in\mathbf{C}$ is the spectral parameter and $m_i^2=\frac{1}{4}|\va_i|^2 m^2 n_i$. For a classical integrable system, the equations of motion are equivalent to the Lax equation
\be\label{Laxeq}
\p \bar{L}-\pb L-\[L,\bar{L}\]=0.
\ee
For later convenience we introduce two new combinations
\be
&E_{+}=\sum_{i=0}^r m_i  e^{\beta \va_i\cd\vp/2}E_{\va_i},\quad E_{-}=\sum_{i=0}^r m_i  e^{\beta \va_i\cd\vp/2}E_{-\va_i}
\ee
satisfying 
\be
\[E_{+},E_{-}\]=-\f{\beta}{2}\n{V}\cd\vH,\hs{3ex}
\[\vH,E_{\pm}\]=\pm\f{2}{\beta}\n{E_{\pm}}
\ee
where $\n{f}$ denotes $\f{\delta f}{\delta \vp}$. Then the Lax connections \eqref{Lax1} can be rewritten as
\be\label{Lax2}
L&=-\frac{\beta}{2}\pp\cd\vH-\la E_{+},\\
\bar{L}&=\frac{\beta}{2}\pbp\cd\vH+\f{1}{\la}E_{-}.
\ee

\subsection{\ttb-deformed theories}
The \ttb-deformed Lagrangian of a $N$-scalar theory is  \cite{closed, LaxGordon}
\be\label{Lagrangian}
\mathcal{L}^{(t)}=\frac{V}{1-t V}+\frac{1}{2 t(1-t V)}(\om-1)
\ee
where
\be\label{Omega}
&\om=\sqrt{1+Y+Z},\quad Y=4t(1-t V)\(\pp\cd\pbp\),\\
&Z=-4 t^2 (1-t V)^2\[\(\pp\cd\pp\)\(\pbp\cd\pbp\)-\(\pp\cd\pbp\)^2\].
\ee
The equations of motion are given by
\be\label{eom}
\vec{A_e}\equiv \p_{\mu}\f{\delta \mathcal{L}^{(t)}}{\delta \p_{\mu}\vp}-\f{\delta \mathcal{L}^{(t)}}{\delta \vp}=0.
\ee
Substituting \eqref{Lagrangian} and \eqref{Omega} into \eqref{eom}, one can get
\be\label{Ae}
\vec{A}_e
=&\p\lb\f{1}{\om}\[\pbp-2t(1-t V)\(\pp\(\pbp\cd\pbp\)-\pbp\(\pp\cd\pbp\)\)\]\rb\\
&+\pb\lb\f{1}{\om}\[\pp-2t(1-t V)\(\pbp\(\pp\cd\pp\)-\pp\(\pp\cd\pbp\)\)\]\rb\\
&-\f{\n{V}}{4\om(1-t V)^2}\[\(\om+1\)^2-Z\].
\ee
Given these equations of motions we propose a simple ansatz for the Lax connection:
\be\label{ansatz1}
&L=-\f{\beta}{2}\vec{a}_1\cd\vH-\la b_1 E_{+}+\f{1}{\la} c_1 E_{-},\\
&\bar{L}=\f{\beta}{2}\vec{a}_2\cd\vH-\la b_2 E_{+}+\f{1}{\la} c_2 E_{-},\\
\ee
where $\vec{a}_1, b_1, c_1, \vec{a}_2, b_2, c_2$ are the functions of $\vp$ and their derivatives, and will be determined by imposing the Lax equation and the equations of motion. Notice that in our ansatz \eqref{ansatz1} the Lax connection depends uniformly on the simple roots $E_{\va}$.
Plugging \eqref{ansatz1} into \eqref{Laxeq} directly gives rise to a set of linear differential equations $\vec{A}_H, A_+',A_-',$ corresponding to the components $\vH$,\,$E_+$ and $E_-$, respectively. In principle $\vec{A}_H, A_+',A_-',$ should vanish separately. However because the terms like $\n E_\pm$ are not uniformly dependent  on the simple roots $E_{\va}$, we would require that the coefficients before the terms like $\n E_\pm$ vanish separately. Consequently we obtain five sets of linear equations
\be\label{Laxeq2}
\left\{
\begin{aligned}
&\vec{A}_H \equiv \p\vec{a}_2+\pb\vec{a}_1-\n{V}(b_1 c_2 -b_2 c_1)=0\\
&A_{+} \equiv -\p b_2+\pb b_1=0\\
&A_{-} \equiv \p c_2+-\pb c_1=0\\
&\vec{A}_{p+} \equiv -\pp b_2+\pbp b_1-(\vec{a}_1 b_2+\vec{a}_2 b_1)=0\\
&\vec{A}_{p-} \equiv \pp c_2-\pbp c_1-(\vec{a}_1 c_2+\vec{a}_2 c_1)=0\\
\end{aligned}
\right.
\ee\par
To solve these equations we make another  assumption that they can be written as linear combinations of the equations of motion, i.e.
\be
\vec{A}_H=f_H\vec{A}_e,\quad A_{+}=\vec{f}_{+}\cd\vec{A}_e,\quad A_{-}=\vec{f}_{-}\cd\vec{A}_e,\quad \vec{A}_{p+}=f_{p+}\vec{A}_e,\quad \vec{A}_{p-}=f_{p-}\vec{A}_e.
\ee
To ensure the equivalence between the Lax equation \eqref{Laxeq2} and the equations of motion \eqref{eom}, there should be no common zero of $f_H, \vec{f}_{+}, \vec{f}_{-}, f_{p+}, f_{p-}$.
Indeed we are making quite strong assumptions here, but we will show that the consistent solution does exist.  

For the undeformed theory, by \eqref{Lax2}, one can find that
\be\label{unde}
f_H=1,\quad f_{p+}=0,\quad f_{p-}=0, 
\ee
and there is no $\vec{f}_{+}, \vec{f}_{-}$ terms. We assume that \eqref{unde} is still true for the deformed theory and observe that if we take
\be\label{assum}
\vec{f}_{+}=-t\pbp,
\ee
then
\be
\vec{f}_{+}\cd\vec{A}_e=-\p\[\f{t}{\om}\(\pbp\cd\pbp\)\]+\pb\[\f{(\om+1)^2-Z}{4\om(1-t V)}\]
\ee
suggesting that we can identify
\be\label{b1b2}
&b_1=\f{(\om+1)^2-Z}{4\om(1-t V)},\quad b_2=\f{t}{\om}\(\pbp\cd\pbp\)
\ee
up to some constants which can be fixed to be zero after considering other equations. Similarly, by taking $\vec{f}_{-}=-t\pp,$ we can read off $c_1$ and $c_2$ 
\be\label{c1c2}
&c_1=\f{t}{\om}\(\pp\cd\pp\),\\
&c_2=\f{(\om+1)^2-Z}{4\om(1-t V)}.\\
\ee
Finally from $\vec{f}_{H}\vec{A}_e$, we fix all the remaining functions in our ansatz 
\be\label{a1a2}
&\vec{a}_1=\f{1}{\om}\[\pp-2t(1-t V)\(\pbp\(\pp\cd\pp\)-\pp\(\pp\cd\pbp\)\)\],\\
&\vec{a}_2=\f{1}{\om}\[\pbp-2t(1-t V)\(\pp\(\pbp\cd\pbp\)-\pbp\(\pp\cd\pbp\)\)\].\\
\ee
Plugging \eqref{b1b2} \eqref{c1c2} and \eqref{a1a2} into \eqref{Laxeq2}, one can check that \eqref{Laxeq2} is indeed equivalent to the equations of motion \eqref{eom}.\par

To summarize,  the Lax connections of the \ttb-deformed (affine) Toda field theories are of the forms \eqref{ansatz1} with the functions being given by \eqref{b1b2} \eqref{c1c2} and \eqref{a1a2}. We want to stress that after we assume \eqref{unde} and \eqref{assum} the solutions  can be read off directly without solving any other equations.

\subsection{Examples}
To compare with the existing results in the literature,  let us consider some specific examples. The first one is the Liouville field theory which corresponds to Toda field theory of  $sl_2$ Lie algebra with the parameters being 
\be
\beta=\frac{1}{2},\ \ \ \ m_0=0,\ \ \ \ m_1=-\frac{\sqrt{\mu}}{2}.
\ee
The undeformed Lagrangian is 
\bea 
\mathcal{L}^{(0)}=\p \phi \bar{\p}\phi-\mu e^\phi.
\eea 
The \ttb-deformed Liouville field theory was studied in \cite{Liouville} where infinite conserved currents were constructed from some ansatz without using the Lax connection. From the discussion in the last subsection, we can  present the  deformed Lax connections explicitly 
\be
L&=-\frac{1}{4}\frac{\p\phi}{\om}H+ \sqrt{\mu}\la B e^{ \phi/2}E_{\alpha_1}- \frac{\sqrt{\mu}}{\la}(\p\phi)^2 C e^{ \phi/2}E_{-\alpha_1},\\
\bar{L}&=\frac{1}{4}\frac{\pb\phi}{\om}H- \frac{\sqrt{\mu}}{\la} B  e^{ \phi/2}E_{-\alpha_1}+\sqrt{\mu}\la (\pb\phi)^2 C e^{ \phi/2}E_{\alpha_1},
\ee
where
\be
&B=\frac{(\om+1)^{2}}{8 \om(1-t V)}, \quad C=\frac{t}{2 \om}\\
&\om=\sqrt{1+4t(1-t V)\(\p\phi\pb\phi\)}. 
\ee
Let the generators of $sl_2$ Lie algebra  be
\be
H=\left(
    \begin{array}{cc}
      1 & 0 \\
      0 & -1 \\
    \end{array}
  \right),\ \
E_{\alpha_1}=\left(
    \begin{array}{cc}
      0 & 1 \\
      0 & 0 \\
    \end{array}
  \right),\ \
E_{-\alpha_1}=\left(
    \begin{array}{cc}
      0 & 0 \\
      1 & 0 \\
    \end{array}
  \right),\ \
\ee
then if we take the undeforming limit, $t \rightarrow 0$, the Lax connections become
\be
L=\left(
    \begin{array}{cc}
      -\frac{1}{4}\p\phi & \frac{\sqrt{\mu}\la }{2} e^{ \phi/2} \\
      0 & \frac{1}{4}\p\phi \\
    \end{array}
  \right),\ \
\bar{L}=\left(
    \begin{array}{cc}
      \frac{1}{4}\pb\phi & 0 \\
      -\frac{\sqrt{\mu} }{2\la} e^{ \phi/2} & -\frac{1}{4}\pb\phi \\
    \end{array}
  \right),\ \
\ee
which are the Lax connections of the Liouville field theory. 

Our next example is the sine-Gordon model which corresponds to  the affine Toda field of affine $sl_2$ algebra with parameters
\be
\beta=\frac{i}{2},\ \, m_0=m_1=-\frac{i}{2},\ \, n_0=n_1=1, \alpha_0=-2,\ \ \alpha_1=2.
\ee
The undeformed Lagrangian is given by
\be
\mathcal{L}^{(0)}=\partial \phi \bar{\partial} \phi -2\cos{\phi }
\ee
By setting
\be
E_{\alpha_{0}}=E_{-\alpha_{1}},\ \ \ \ E_{-\alpha_{0}}=E_{\alpha_{1}},
\ee
we find that the deformed Lax connections are
\be
L&=-\frac{i}{4}\frac{\p\phi}{\om}H+\( i\la B e^{i\phi/2}+\frac{1}{i\la}(\p\phi)^2 C e^{-i \phi/2}\) E_{\alpha_1} + \(i\la B e^{-i\phi/2}+\frac{1}{i\la}(\p\phi)^2 C e^{i \phi/2}\)E_{-\alpha_1},\\
\bar{L}&=\ \ \ \frac{i}{4}\frac{\pb\phi}{\om}H+\( \frac{1}{i\la}B e^{-i\phi/2}+i\la(\pb\phi)^2 C e^{i \phi/2}\) E_{\alpha_1} + \( \frac{1}{i\la}B e^{i\phi/2}+i\la(\pb\phi)^2 C e^{-i \phi/2}\) E_{-\alpha_1}\\
\ee
which are the same as the ones found in \cite{LaxGordon}.
In the undeforming limit, $t \rightarrow 0$, the Lax connections reduce to the Lax connections of sine-Gordon model
\be
L=\left(
    \begin{array}{cc}
      -\frac{i}{4}\p\phi & \frac{i \la }{2} e^{ i\phi/2} \\
      \frac{i \la }{2} e^{-i\phi/2}  & \frac{i}{4}\p\phi \\
    \end{array}
  \right),\ \
\bar{L}=\left(
    \begin{array}{cc}
      \frac{i}{4}\pb\phi & \frac{1}{2 i\la} e^{-i \phi/2} \\
      \frac{1}{2 i\la} e^{i \phi/2} & -\frac{i}{4}\pb\phi \\
    \end{array}
  \right).\ \
\ee

\section{ Principal chiral model}
In this section we consider the principal chiral model (PCM) which is an integrable sigma model. The \ttb-deformed Lagrangian of PCM has been obtained in \cite{closed,LaxGordon}. We will use a similar strategy used in the last section to derive the deformed Lax connection.

\subsection{Undeformed theory}
A principal chiral model (PCM) is a field theory whose field takes values in some Lie group manifold. Its action is\footnote{In the section, we consider the theory in the flat space and take the Euclidean signature, that is, $g^{\mu\nu}=diag(1,1)$. The coordinate is $(x^0, x^1)$ and the Levi-Civita symbol is $\ep^{0,1}=-\ep^{1,0}=1$.}
\be
S_0=\int dx^2 g^{\mu\nu} \Tr\(g^{-1}\p_{\mu}g g^{-1}\p_{\nu}g\),\quad g\in G.
\ee
Usually the Lie group is chosen to be semisimple but we will leave it to be arbitrary since our interest is on integrability. The model has symmetry group $G_L\times G_R$.
The equation of motion of PCM is just
\be
\p^{\mu} \(g^{-1}\p_{\mu}g\)=0,
\ee
which is equivalent to the current conservation equation 
\bea\label{conservation}
\p_\mu j^\mu=0,\quad j^\mu \equiv g^{-1}\p^{\mu}g. 
\eea 
Here $j^\mu$ is the conserved current corresponding to the $G_R$ symmetry. In addition to \eqref{conservation}, the conserved current also satisfies the flatness condition:
\be\label{flat}
\p_0 j_1-\p_1 j_0 = -\[j_0, j_1\].
\ee
The equations \eqref{conservation} and \eqref{flat} are equivalent to the Lax equation with the Lax connections
\be\label{laxu}
&L_0=-\f{1}{\la^2+1}(\la j_1+ j_0),\\
&L_1=-\f{1}{\la^2+1}(-\la j_0+j_1),\\
\ee
where $\la \in \mathbf{C}$ is the spectral parameter.

\subsection{\ttb-deformed theory}
The \ttb-deformed Lagrangian of PCM is given by \cite{closed}
\be
\mathcal{L}^{(t)}_{PCM}=\f{1}{2 t} \(-1+\omp\),
\ee
where
\be
\omp&=\sqrt{1+4 t \Tr\(g^{-1}\p_{\mu}g g^{-1}\p_{\mu}g\)+8 t^2 \ep^{\mu\nu}\ep^{\rho\si}\Tr\(g^{-1}\p_{\mu}g g^{-1}\p_{\rho}g\)\Tr\(g^{-1}\p_{\nu}g g^{-1}\p_{\si}g\)}\\
&=\sqrt{1+4 t \Tr\(j_{\mu}j^{\mu}\)+8 t^2 \ep^{\mu\nu}\ep^{\rho\si}\Tr\(j_{\mu}j_{\rho}\)\Tr\(j_{\nu}j_{\si}\)}.\\
\ee
The equation of motion $A_{ePCM}=0$, can also be cast into a form of conservation law:
\be
A_{ePCM}&\equiv \p_{\mu}\f{\delta \mathcal{L}^{(t)}_{PCM}}{\delta \p_{\mu}\vp}-\f{\delta \mathcal{L}^{(t)}_{PCM}}{\delta \vp}=2(\p_{\mu} J^{\mu}) g^{-1}
\ee
Here the conserved current $J^{\mu}$ is defined as
\be\label{Jmu}
J^{\mu}=\f{1}{\omp}\(j^{\mu}+4 t \ep^{\mu\nu}\ep^{\rho\si}j_{\rho}\Tr\(j_{\nu}j_{\si}\)\),
\ee
which satisfies the following useful identities
\be\label{identities}
&[J_0, J_1]=[j_0, j_1],\\
&[J_0, j_0]=\f{1}{\omp}4 t \Tr\(j_{0}j_{1}\)[j_0, j_1],\\
&[J_0, j_1]=\f{1}{\omp}\(1+4 t \Tr\(j_{1}j_{1}\)\)[j_0, j_1],\\
&[J_1, j_0]=-\f{1}{\omp}\(1+4 t \Tr\(j_{0}j_{0}\)\)[j_0, j_1],\\
&[J_1, j_1]=-\f{1}{\omp}4 t \Tr\(j_{0}j_{1}\)[j_0, j_1].\\
\ee\par
Notice that the current $j_\mu$ still satisfies the flatness condition \eqref{flat} so the reasonable ansatz for the Lax connections could be that they are the linear combination of the new conserved current \eqref{Jmu} and  $j_\mu$:
\be\label{L0L1}
&L_0=a_0 J_1+b_0 j_0+c_0 j_1,\\
&L_1=a_1 J_0+b_1 j_0+c_1 j_1, \\
\ee
where $a_0, a_1, b_0, b_1, c_0, c_1$ are some constants to be determined. Again we assume that the Lax equation is linearly dependent on the equation of motion:

\be\label{AL1}
A_{LPCM}\equiv& \p_0 L_1-\p_1 L_0 -\[L_0, L_1\],\quad A_{LPCM}=A_{ePCM} \cd f_{PCM}\\
\ee
For the undeformed theory, using \eqref{laxu} and the definition of $A_{LPCM}$ and $A_{ePCM}$, we can get
\be\label{Unde2}
f_{PCM}=\f{1}{2}\f{\la}{\la^2+1}g
\ee 
Assuming that \eqref{Unde2} is still true in the deformed case, we end up with
\be\label{AL2}
A_{LPCM}=\f{\la}{\la^2+1}\p_{\mu} J^{\mu}. 
\ee
Plugging \eqref{L0L1} into \eqref{AL1} and matching it with \eqref{AL2} we can read off
\be
a_0=-a_1=-\f{\la}{\la^2+1},\quad b_0=c_1=-\f{1}{\la^2+1},\quad b_1=c_0=0, 
\ee
where we have used the identity, $\p_0 j_1-\p_1 j_0 = -\[j_0, j_1\]$.\par

In summary, the Lax connection of the \ttb-deformed PCM is given by
\be
&L_0=-\f{1}{\la^2+1}(\la J_1+ j_0),\\
&L_1=-\f{1}{\la^2+1}(-\la J_0+j_1),\\
\ee
where $J_{\mu}$ has been defined by \eqref{Jmu}. This result is expected considering the identities \eqref{identities}. Given the Lax connection, we can define the monodromy matrix as the holonomy along a constant time slice
\bea 
M(x^0;\lambda)=\mathcal{P}\exp\(\int_{-\infty}^{\infty} dx^1\, L_1(x^0,x^1,\lambda)\).
\eea 
The set of (non-local) infinite conserved charges can be generated by expanding the monodromy matrix with respect to the spectral parameter as
\bea \label{conserved}
&&M(\lambda)=\exp \(\sum_{n=1}^\infty \frac{Q_n}{z^n}\)\\
&&\quad =1+\frac{1}{\la} \int_{-\infty}^{+\infty} dx^1 J_0-\frac{1}{\la^2}\(\int_{-\infty}^{+\infty}dx^1 \, j_1- \int_{-\infty}^{+\infty}dx^1 \int_{-\infty}^{x^1}dy^1 J_0(x)J_0(y)\)+\mathcal{O}(\frac{1}{\lambda^3}).\nonumber
\eea 
For the undeformed PCM these non-local charges span the classical Yangian algebra \cite{Yangian}. Under the \ttb-deformation, the algebra gets deformed in a very complicated way.

\section{Lax connections from dynamical coordinate transformation}

The solvability of \ttb-deformation can be understood in various ways. From the point of view of integrability\footnote{Here specifically by integrability we mean there exist infinite conserved charges.}, the most transparent approach  is to realize the \ttb-deformation as dynamical coordinate transformation.   It was shown in \cite{CT1,CT2,cdd}, the \ttb deformation can be interpreted as a space-time deformation. In Euclidean signature  the deformed and undeformed space-time are related via the following (state dependent or dynamical) coordinate transformation
\bea 
&&dx^\mu=\left(\delta^\mu_{~\nu}+t\, \tilde{T}^\mu_{~\nu}(\mathbf{y})\right)dy^\nu,\quad \mathbf{y}=(y^1,y^2),\\
&&dy^{\mu}=\left(\delta^\mu_{~\nu}+t \,  (\tilde{T}^{(\tau)})^\mu_{~\nu}   (\mathbf{x})\right)dx^\nu,\quad \mathbf{x}=(x^1,x^2),
\eea 
with $\tilde {T}^\mu_{~\nu}=-\epsilon^\mu_{~\rho}\epsilon^\sigma_{~\nu}T^\rho_{~\sigma}$ and $(\tilde {T}^{(\tau)})^\mu_{~\nu}=-\epsilon^\mu_{~\rho}\epsilon^\sigma_{~\nu}(T^{\tau})^\rho_{~\sigma}$, where $T=T^{(0)}$ and $T^{(\tau)}$ are the undeformed and deformed stress-energy tensor in the coordinates $\mathbf{y}$ and $\mathbf{x}$, respectively. Using this map we can obtain the solutions of the deformed equation of motions as
\bea 
\phi^{(\tau)}(\mathbf{x})=\phi^{(0)}(\mathbf{y}(\mathbf{x})).
\eea

Apart from the solutions of equation of motions, the deformed conserved currents can also be obtained from the undeformed ones by using the above coordinate transformations\cite{CT2}. First let us switch to complex coordinates defined by
\bea 
&&z=x^1+i x^2,\quad \zb=x^1-i x^2,\\
&&w=y^1+i y^2,\quad \wb=y^1-i y^2. 
\eea
Starting from the 1-forms in the $\mathbf{w}$ coordinates 
\bea \label{ConJ}
\mathcal{J}_k&=&T_{k+1}(\mathbf{w})dw+\Theta_{k-1}(\mathbf{w})d\wb,\nn
\bar{\mathcal{J}}_k&=&\bar{T}_{k+1}(\mathbf{w})d\wb+\bar{\Theta}_{k-1}(\mathbf{w})dw,
\eea
where $T_{k+1}$, $\Theta_{k-1}$ and their complex conjugates are the higher conserved currents of underformed theory. Under the change of coordinates, we have \bea \label{Cc}
\begin{pmatrix}
dw\\
d\wb
\end{pmatrix}=\mathcal{J}^T \begin{pmatrix}
dz\\
d\zb
\end{pmatrix},\quad \mathcal{J}=\begin{pmatrix}
\p w&\p \bar{w}\\
\bar{\p}w& \bar{\p}\wb
\end{pmatrix}.
\eea 
where $\p$ and $\bar{\p}$ denote the derivative with respect to $z$ and $\bar{z}$, respectively. Now the Jacobian is of the form  
\bea \label{Cc1}
\mathcal{J}=\frac{1}{\Delta(\mathbf{w})} \begin{pmatrix}
1+2 t \Theta_0(\mathbf{w}) & -2 tT_2(\mathbf{w}) \\
-2 t \bar{T}_2(\mathbf{w}) & 1+2 t \bar{\Theta}_0(\mathbf{w})
\end{pmatrix}
\eea 
with
\bea 
\Delta(\mathbf{w})=(1+2 t \Theta_0(\mathbf{w}))(1+2 t \bar{\Theta}_0(\mathbf{w}))-4 t^2 T_2(\mathbf{w})\bar{T}_2(\mathbf{w}).
\eea
Substituting \eqref{Cc} and \eqref{Cc1} into \eqref{ConJ} one can read off the components of the currents in the $\mathbf{z}$ coordinates:
\bea \label{TransformCurrent}
T_{k+1}(\mathbf{z},t)&=&\frac{T_{k+1}(\mathbf{w}(\mathbf{z}))+2 t (T_{k+1}(\mathbf{w}(\mathbf{z}))\Theta_0(\mathbf{w}(\mathbf{z}))-\Theta_{k-1}(\mathbf{w}(\mathbf{z})) T_2(\mathbf{w}(\mathbf{z})))}{\Delta(\mathbf{w}(\mathbf{z}))},\\
\Theta_{k-1}(\mathbf{z},t)&=&\frac{\Theta_{k-1}(\mathbf{w}(\mathbf{z}))+2 t (\Theta_{k-1}(\mathbf{w}(\mathbf{z}))\bar{\Theta}_0(\mathbf{w}(\mathbf{z}))-T_{k+1}(\mathbf{w}(\mathbf{z}))\bar{T}_2(\mathbf{w}(\mathbf{z})))}{\Delta(\mathbf{w}(\mathbf{z}))}.
\eea 

In a similar way, we can read the Lax connection of the deformed model. If 
 the Lax connection of the undeformed model is
\bea 
L(w,\bar{w})=\mathcal{L} \,dw+\bar{\mathcal{L}}\,d\bar{w}
\eea 
one can expect the deformed Lax pair should be given by
\bea\label{LaxAn}
L&=&\mathcal{L}(z,\zb)dz+\bar{\mathcal{L}}(z,\zb)d\zb \nn
&=&\mathcal{L}(w,\wb)\left(\frac{\p w}{\p z}dz+\frac{\p w}{\p \zb}d\zb\right)+\bar{\mathcal{L}}(w,\wb)\,\left(\frac{\p \wb}{\p z}dz+\frac{\p \wb}{\p \zb}d\zb\right), 
\eea 
which leads to the transformation law on the Lax connections:
\bea \label{LaxTransform}
\mathcal{L}(\mathbf{z},t)&=&\frac{\mathcal{L}_w(\mathbf{w}(\mathbf{z}))+2t(\mathcal{L}_w(\mathbf{w}(\mathbf{z}))\Theta_0(\mathbf{w}(\mathbf{z}))-\mathcal{L}_{\wb}(\mathbf{w}(\mathbf{z}))T_2(\mathbf{w}(\mathbf{z})))}{\Delta(\mathbf{w}(\mathbf{z}))},\\
\bar{\mathcal{L}}(\mathbf{z},t)&=&\frac{\mathcal{L}_{\wb}(\mathbf{w}(\mathbf{z}))+2t(\mathcal{L}_{\wb}(\mathbf{w}(\mathbf{z}))\bar{\Theta}_0(\mathbf{w}(\mathbf{z}))-\mathcal{L}_{w}(\mathbf{w}(\mathbf{z}))\bar{T}_2(\mathbf{w}(\mathbf{z})))}{\Delta(\mathbf{w}(\mathbf{z}))}.
\eea 
In the following, we will check the above relations in a free scalar theory and the sine-Gordon model whose deformed Lax pairs are explicitly given in the literature \cite{LaxGordon}. Moreover, we will try to reproduce the Lax connections of affine Toda field theory and PCM which we found in previous sections. 

\bigskip


\subsection{Free scalar}
Consider the free scalar with Lagrangian
\bea \label{FreeS}
L(\mathbf{w})=\p_w \phi \p_{\wb} \phi.
\eea 
The model is integrable with trivial Lax pair
\bea 
\mathcal{L}_w=\p_w \phi,\quad \mathcal{L}_{\wb}=-\p_{\wb} \phi
\eea 
such that the Lax equation
\bea 
\p_{\wb}\mathcal{L}_w-\p_w \mathcal{L}_{\wb}=2\p_w \p_{\bar{w}}\phi=0
\eea 
coinciding with the equation of motion. The stress-energy tensor is simply
\bea 
T_2(\mathbf{w})=-\frac{1}{2} (\p_w \phi)^2,\quad \Theta_0(\mathbf{w})=0,\quad \Delta=1-4t^2 T_2(\mathbf{w})\bar{T}_2(\mathbf{w}),
\eea 
which leads to the following transformation
\bea 
&&\p_w \phi=\p \phi-\frac{1}{4\tau}\left(\frac{-1+\om}{\bar{\p}\phi}\right)^2 \bar{\p}\phi,\quad \p_{\bar{w}}\phi=\bar{\p} \phi-\frac{1}{4 t}\left(\frac{-1+\om}{{\p}\phi}\right)^2\p {\phi},\eea
with $\om=\sqrt{1+4 t\p\phi\bar{\p}\phi}.$ 
Therefore the deformed Lax connection is given by
\bea \label{LaxTransformfree}
&&\mathcal{L}(\mathbf{z},\tau)=\frac{\p_w \phi(\mathbf{w}(\mathbf{z}))+2 t \p_{\wb}\phi(\mathbf{w}(\mathbf{z}))T_2(\mathbf{w}(\mathbf{z}))}{1-4t^2 T_2(\mathbf{w}(\mathbf{z}))\bar{T}_2(\mathbf{w}(\mathbf{z}))}=\frac{\p\phi}{\om},\\
&&\bar{\mathcal{L}}(\mathbf{z},\tau)=\frac{-\p_{\wb} \phi(\mathbf{w}(\mathbf{z}))-2t \p_{w}\phi(\mathbf{w}(\mathbf{z}))\bar{T}_2(\mathbf{w}(\mathbf{z}))}{1-4t^2 T_2(\mathbf{w}(\mathbf{z}))\bar{T}_2(\mathbf{w}(\mathbf{z}))}=-\frac{\bar{\p}\phi}{\om}.
\eea 
Indeed the Lax equation matches the equation of motion of the \ttb-deformed free scalar:
\bea 
\p\left(\frac{\bar{\p}\phi}{\om}\right)+\bar{\p}\left(\frac{\p\phi}{\om}\right)=0.
\eea 

\subsection{Sine-Gordon model}
Next, we turn to the \ttb-deformed sine-Gordon model, whose Lax pair has been given in \cite{LaxGordon}. As a first step, we need to find the Jacobian \eqref{Cc1} which is determined by the stress-energy tensor in the $\mathbf{w}$ space-time. The Lagrangian of the sine-Gordon model is given by adding the potential\footnote{Here we have chosen the convention used in \cite{LaxGordon} in order to make the comparison.}
\bea
V=4 \sin^2(\frac{\phi}{2})
\eea 
to the free scalar Lagrangian \eqref{FreeS}. From the standard procedure one can find the expression of the stress-energy tensor
\bea 
T_2(\mathbf{w})=-\frac{1}{2}(\p_w \phi)^2,\quad \bar{T}_2(\mathbf{w})=-\frac{1}{2}(\p_{\wb} \phi)^2,\quad \Theta_0(\mathbf{w})=-2\sin^2(\frac{\phi}{2})
\eea
which leads to the following transformations
\bea \label{1ScalarTran}
\p_w \phi=\frac{-1+\om}{2t \bar{\p}\phi},\quad \p_{\wb} \phi=\frac{-1+\om}{2t \p\phi},\quad \om=\sqrt{1+4t(1-tV)\p \phi \bar{\p}\phi}.
\eea 
Recall that the undeformed Lax connection is 
\bea 
&&\mathcal{L}_w=-\frac{i}{4}\p_w\phi H+\frac{\lambda}{2}e^{i\frac{\phi}{2}}E_++\frac{\lambda}{2}e^{-i\frac{\phi}{2}}E_-,\\
&&\mathcal{L}_{\wb}=\frac{i}{4}\p_{\wb}\phi H+\frac{1}{2\lambda}e^{-i\frac{\phi}{2}}E_++\frac{1}{2\lambda}e^{i\frac{\phi}{2}}E_-.
\eea 
The deformed Lax connection can be expanded with respect these three generators as
\bea \label{decompose}
\mathcal{L}(\mathbf{z},t)=\mathcal{L}^0 H+\mathcal{L}^+ E_++\mathcal{L}^- E_-,\quad \bar{\mathcal{L}}(\mathbf{z},\tau)=\bar{\mathcal{L}}^0 H+\bar{\mathcal{L}}^+ E_++\bar{\mathcal{L}}^- E_-.
\eea
From the transformation \eqref{LaxTransform}, we have the deformed Lax connections
\bea 
\mathcal{L}^0=-\frac{i \p \phi}{4 \om},\quad &&\bar{\mathcal{L}}^0=\frac{i \bar{\p} \phi}{4 \om},\\
\mathcal{L}^+=\frac{e^{-i\frac{\phi}{2}}}{\lambda}\frac{(\p\phi)^2t}{2\om}+\lambda e^{i\frac{\phi}{2}}\frac{(\om+1)^2}{8\om(1-t V)},\quad &&\bar{\mathcal{L}}^+=\lambda{e^{i\frac{\phi}{2}}}{}\frac{(\bar{\p}\phi)^2t}{2\om}+\frac{ e^{-i\frac{\phi}{2}}}{\lambda}\frac{(\om+1)^2}{8\om(1-t V)},\nn
\mathcal{L}^-=\frac{e^{i\frac{\phi}{2}}}{\lambda}\frac{(\p\phi)^2t}{2\om}+\lambda e^{-i\frac{\phi}{2}}\frac{(\om+1)^2}{8\om(1-t V)},\quad &&\bar{\mathcal{L}}^-=\lambda{e^{-i\frac{\phi}{2}}}{}\frac{(\bar{\p}\phi)^2t}{2\om}+\frac{ e^{i\frac{\phi}{2}}}{\lambda}\frac{(\om+1)^2}{8\om(1-tV)},\nonumber
\eea
which coincide with the ones found in \cite{LaxGordon}.

\subsection{Liouville field theory}
The Lagrangian of the classical Liouville field theory is 
\bea 
\mathcal{L}(\mathbf{w})=\p_w \phi \p_{\wb}\phi-\mu e^\phi,\quad V=-\mu e^{\phi}
\eea
with the Lax connection
\bea 
&&\mathcal{L}_w=-\p_w\phi H+2\lambda\sqrt{\mu}e^{\frac{\phi}{2}}E_+,\quad \mathcal{L}_{\wb}={\p_{\wb}}\phi H-\frac{1}{2\lambda}\sqrt{\mu}e^{\frac{\phi}{2}}E_-.
\eea 
The field transformation is also given by \eqref{1ScalarTran}. Decomposing the Lax connection as \eqref{decompose} again we find
\bea 
\mathcal{L}^0=-\frac{ \p \phi}{ \om},\quad &&\bar{\mathcal{L}}^0=\frac{\bar{\p} \phi}{ \om},\\
\mathcal{L}^+=\frac{\lambda\sqrt{\mu}e^{\frac{\phi}{2}}(1+\om)^2}{2\om(1-tV)},\quad &&\bar{\mathcal{L}}^+=\frac{2t\la\sqrt{\mu} (\pb \phi)^2 }{\om},\nn
\mathcal{L}^-=-\frac{t\sqrt{\mu} e^{\frac{\phi}{2}}(\p\phi)^2 }{2\la \om},\quad&& \bar{\mathcal{L}}^-=-\frac{\sqrt{\mu}e^{\frac{\phi}{2}}(1+\om)^2}{8\la \om(1-tV)}.\nonumber
\eea 
They differ from the ones in (25) up to numerical factors, due to different conventions. 

With these deformed Lax connection one can derive infinite conserved charges. On the other hand, the (anti)-holomorphic currents are simply given by taking powers of the modified traceless stress-energy tensor
\bea 
T_{2n}=\left((\p_w\phi)^2-2\p_w^2 \phi \right)^n,\quad \bar{T}_{2n}=\left((\bar{\p}_w\phi)^2-2\bar{\p}_w^2 \phi \right)^n.
\eea 
From \eqref{1ScalarTran} and \eqref{TransformCurrent} one can read the deformed currents
\bea 
&&T_{2n}(\mathbf{z})=-\frac{\om+(2t(1-tV)\p\phi \bar{\p}\phi+1)}{2\om(1-\tau V)}T_{2n}(\mathbf{w}(\mathbf{z})),\\
&&\Theta_{2n}(\mathbf{z})=\frac{t(\bar{\p}\phi)^2}{\om}T_{2n}(\mathbf{w}(\mathbf{z})).\nonumber
\eea 
The explicit expressions of these currents have been  derived in \cite{Liouville} using a different method.

\subsection{$N$ bosonic scalars with arbitrary potential}
To construct the deformed Lax connections for the (affine) Tode field theories,  let us first consider the $N$ free scalars with arbitrary potential \cite{closed, LaxGordon}
\bea 
\mathcal{L}_N=\sum_i ^N \p_w \phi_i \p_{\wb}\phi_i+V(\phi_i).
\eea 
From the relationships
\bea \label{xy}
\frac{\p x^1}{\p y^1}=1+t T^2_{~2}(\mathbf{y}),\quad \frac{\p x^2}{\p y^2}=1+t T^1_{~1}(\mathbf{y}),\quad \frac{\p x^1}{\p y^2}=\frac{\p x^2}{\p y^1}-=-t T^1_{~2}(\mathbf{y}),
\eea 
we can compute the inverse of the Jacobian
\bea 
\mathcal{J}_N^{-1}=\begin{pmatrix}
\p _w z&\p_w \bar{z}\\
\p_{\bar{w}}z&\p_{\wb}\zb 
\end{pmatrix}=\begin{pmatrix}
1-t V&-t \sum_i (\p_w \phi_i)^2\\
-t \sum_i (\p_{\wb}\phi_i)^2&1-t V
\end{pmatrix}.
\eea 
The main technical difficulty of this method is to solve $\p_w \phi_i$ and $\p_{\wb}\phi_i$ from
\bea \label{ChainRule}
\begin{pmatrix}
\p_w \phi_i \\ \p_{\wb} \phi_i 
\end{pmatrix}=\mathcal{J}_N^{-1}\begin{pmatrix}
\p \phi_i \\ \bar{\p} \phi_i 
\end{pmatrix}
\eea 
in terms of $\p\phi_i$ and $\bar{\p}\phi_i$. For this particular example we find the following solution
\bea 
&&\p_w \phi_i=\frac{1}{2t}\frac{\bar{\p}\phi_i(-1+\om)+\tilde{t} \frac{\p B}{\p \p\phi_i}}{\bar{{K}}},\quad \p_{\wb}\phi=\frac{1}{2t}\frac{{\p}\phi_i(-1+\om)+\tilde{t} \frac{\p B}{\p \bar{\p}\phi_i}}{{{K}}},\eea
with 
\bea
&&\tilde{t}=t(1-t V),\quad \om=\sqrt{1+4\tilde{t}(\mathcal{L}^{(0)}-\tilde{t}B)},\\
&&\mathcal{L}^{(0)}=\sum_{i=1}^N \p\phi_i \bar{\p}\phi_i,\quad B=\sum_{i=1}^N (\p\phi_i)^2\sum_{j=1}^N(\bar{\p}\phi_j)^2-\left(\sum_{i=1}^N \p\phi_i\bar{\p}\phi_i\right)^2,\\
&&K=\sum_i^N (\p \phi_i)^2,\quad \bar{K}=\sum_i^N (\bar{\p}\phi_i)^2.
\eea 
 On the other hand the stress-energy tensor is given by
\bea 
&&K_w=\sum_i^N (\p_w \phi_i)^2,\quad \bar{K}_{\wb}=\sum_i^N (\p_{\wb}\phi_i)^2,\\
&&T_2=-\frac{1}{2}K_w,\quad \bar{T}_2=-\frac{1}{2}\bar{K}_{\wb},\quad \Theta_0=-\frac{1}{2}V. 
\eea 
Therefore the deformed Lax connection is directly given by \eqref{LaxTransform}
\bea \label{Laxdeform}
\mathcal{L}=\frac{(1-\tau V)\mathcal{L}_{w}+\tau K_w \mathcal{L}_{\wb}}{(1-\tau V)^2-\tau^2K_w\bar{K}_{\wb}},\quad \bar{\mathcal{L}}=\frac{(1-\tau V)\mathcal{L}_{\wb}+\tau \bar{K}_{\wb} \mathcal{L}_{w}}{(1-\tau V)^2-\tau^2K_w\bar{K}_{\wb}}. 
\eea 
Using the identities \eqref{ChainRule} we can find the relation between $K_w$, $\bar{K}_{\wb}$ and $K$ and $\bar{K}$
\bea 
&& K_w=(1-\tau V)^2 K+\tau^2K_{w}^2\bar{K}-2\tau K_w(1-\tau V)\mathcal{L}^{(0)},\\
&& \bar{K}_{\wb}=(1-\tau V)^2\bar{K}+\tau^2K_{\wb}^2 K-2\tau \bar{K}_{\wb}(1-\tau V)\mathcal{L}^{(0)}.
\eea 
These are quadratic equations, whose  solutions are\footnote{There are two branches of solutions, here we only keep the one which is consistent with our results in previous section. The other branch gives equivalent result up to a gauge transformation.}
\bea \label{KKbar}
&&K_w=\frac{2\tilde{t}\mathcal{L}^{(0)}+1-\om}{2t^2 \bar{K}},\quad \bar{K}_{\wb}=\frac{2\tilde{t}\mathcal{L}^{(0)}+1- \om}{2\tau^2 {K}},
\eea 
where we used the identity
\bea 
B=K\bar{K}-\mathcal{L}^{(0)}\mathcal{L}^{(0)}.
\eea 
Substituting \eqref{KKbar} into \eqref{Laxdeform} gives
\bea 
&&\mathcal{L}=-\frac{\om+ (2\tilde{t}\mathcal{L}^{(0)}+1)}{2\om(1-t V)}\mathcal{L}_w-\frac{t K}{\om}\mathcal{L}_{\wb},\\&&\bar{\mathcal{L}}=-\frac{\om+ (2\tilde{t}\mathcal{L}^{(0)}+1)}{2\om(1-t V)}\mathcal{L}_{\wb}-\frac{t \bar{K}}{\om}\mathcal{L}_{w}.\label{NscalarP}
\eea 

For the affine Toda theories, whose Lax connections are known, it is straightforward to read the deformed Lax connection from \eqref{NscalarP}. They turn out to be in match with the ones  we derived previously in section 2.2.  Furthermore, we can  use the relations to derive the deformed Lax connection of PCM if we make the following identification 
\bea 
j_\mu=j^i_\mu T_i,\quad j^i_\mu \equiv \p_\mu \phi_i
\eea 
where $T_i$ are the generators of the Lie algebra with the Killing metric $\text{Tr}(T_iT_j)=\delta_{ij}$.

\subsection{Nonlinear Schr\"{o}dinger model}
As our last example, let us consider the \ttb-deformed nonlinear Schr\"{o}dinger model which is a non-relativistic complex field theory. The \ttb-deformed Lagrangian was recently derived in \cite{Non1,Non2,Non3}. Here we derive the deformed Lax connection from the dynamical coordinate transformation.\par
For the undeformed model, the Lagrangian  is 
\bea 
\mathcal{L}_{NS}(y_1,y_2)=\frac{i}{2}\(\qb \p_{y_1} q-q \p_{y_1} \qb \)-\frac{\p_{y_2} q \p_{y_2} \qb}{2m}-g |q\qb|^2,
\eea 
which has the following equations of motion 
\bea \label{NLST}
&&-i\p_{y_1} q=\frac{1}{2m}\p^2_{y_2}q-2g q^2\qb,\quad i\p_{y_1} \qb=\frac{1}{2m}\p^2_{y_2}\qb-2g q\qb^2,\eea
and the stress-energy tensor
\bea
&&T_{y_2y_2}=-\frac{1}{m}\p_{y_2}q \p_{y_2}\qb-\mathcal{L}_{NS}(y_1,y_2),\quad T_{y_2y_1}=-\frac{1}{2m}\(\p_{y_2}\qb\p_{y_1}q+\p_{y_2}q\p_{y_1}\qb\),\\
&&T_{y_1y_2}=\frac{i}{2}\(\qb\p_{y_2}q-q\p_{y_2}\qb\),\quad T_{y_1y_1}=\frac{i}{2}\(\qb\p_{y_1}q-q\p_{y_1}\qb\)-\mathcal{L}_{NS}(y_1,y_2).
\eea 
The corresponding Lax connection is
\bea 
&&U_{y_2}=-i\la \sigma_3+i \sqrt{2gm}Q,\\
&&V_{y_1}=-\frac{i\la^2}{m}\sigma_3+i\sqrt{\frac{2g}{m}}\lambda Q+\sqrt{\frac{g}{2m}}\p_{y_2}Q\sigma_3+igQ^2\sigma_3,
\eea 
where
\bea 
\sigma_3=\begin{pmatrix}
1&0\\
0&-1
\end{pmatrix},\quad Q=\begin{pmatrix}
0&q\\
-\qb&0
\end{pmatrix}.
\eea 
Solving \eqref{xy} one can find the following rules of transformation \cite{Non3}:
\bea \label{NSLTransform}
&&\p_{y_1}q=\frac{2m(B-S)\p_{x_1}\qb+2\tilde{t}\qA C}{2t \qA^2},\quad \p_{y_2}q=\frac{2m(B-S)}{2t \qA},\\
&&\p_{y_1}\qb=\frac{2m(B-S)\p_{x_1}q-2\tilde{t}A C}{2t A^2},\quad \p_{y_2}\qb=\frac{2m(B-S)}{2t A},\nonumber
\eea
where we have defined 
\bea 
&&\tilde{t}=t(1+tV),\quad C=\p_{x_2}\qb \p_{x_1}q-\p_{x_1}\qb \p_{x_2}q,\quad B=1+\frac{it}{2}(\qb \p_{x_1}q-q\p_{x_1}\qb), \\
&&A=\p_{x_2}q+\frac{it}{2}q C,\quad \qA=\p_{x_2}\qb+\frac{it}{2}\qb C,\quad S=\sqrt{B^2-\frac{2\tilde{t}}{m}A\qA}.
\eea
Substituting \eqref{NLST} and \eqref{NSLTransform} into \eqref{LaxTransform}, and after some manipulations we end up with final results of the deformed Lax connection
\bea 
\begin{pmatrix}
V_{x_1}\\
U_{x_2}
\end{pmatrix}=\begin{pmatrix}
J_{11}&J_{12}\\
J_{21}&J_{22}
\end{pmatrix}\begin{pmatrix}
V_{y_1}\\
U_{y_2}
\end{pmatrix}
\eea 
where
\bea 
&&J_{11}=\frac{tB(B+S)}{2S\tilde{t}},\quad J_{12}=-\frac{t(A\p_{x_1}\qb+\qA\p_{x_1}q)}{2mS},\\
&&J_{21}=\frac{it^2(B+S)(\qb \p_{x_2}q-q \p_{x_2}\qb)}{4S \tilde{t}},\quad J_{22}=\frac{2t A\qA}{2mS(B-S)}-\frac{t(A\p_{x_2}\qb+\qA\p_{x_2}q)}{2mS}.\nonumber
\eea 

\section{Conclusion}

In this work we constructed the Lax connections of several  \ttb-deformed integrable models in two different ways, and found consistent picture. The first way is based on  proper ansatz, which assumes that the Lax equation is linearly dependent on the equation of motion. In the discussion, we also assumed that some proportional functions or parameters are invariant under the deformation. We obtained the Lax connections for the affine Toda theories and the principal chiral model. The method  is  suggestive, but its potential is not clear to us.  

The other way relies on the dynamical coordinate transformation between the \ttb-deformed theory and its ancestor.  The method is systematic but maybe difficult to implement  in some models due to the complexity of the dynamical coordinate transformation. We showed the power of the coordinate transformation in several models, including the free scalar theory, sine-Gordon model, Liouville field theory, $N$-scalar theory and non-linear Schr\"{o}dinger model. 

We want to stress that the dynamical coordinate transformation is not a diffeomorphism. Because the coordinate transformation depends on the dynamical fields, the inverse of the transformation could not be obtained in a closed form. Actually we  tried to derive the Lax connection of \ttb-deformed KdV equation. In this case the coordinate transformation depends on the higher order derivatives so the closed form of the inverse of the transformation is unlikely to exist. It is interesting to investigate the effectiveness of the coordinate transformation in other models, for example, the fermionic ones \cite{Fermi}. Besides the two methods discussed in this work, it would be interesting to study the Lax connection from other aspects on \ttb-deformation, say from the light--cone gauge approach in \cite{aspect4}.\par

Given the explicit form of the Lax connection, there are various of applications. The first one is to construct infinite conserved charges as we show for the PCM. The expression \eqref{conserved} indicates the conserved charges get deformed in a very complicated way and it would be very interesting to study how the algebra is deformed. The other application is to construct the solitonic surfaces following \cite{CT1}. Most importantly we hope our construction of Lax connection can shed light on the quantization of the \ttb-deformation.

\section*{Acknowledgements}\noindent
We would like to thank Huajia Wang for valuable discussions.
The work is in part supported by NSFC Grant  No. 11735001. JT is also supported by the UCAS program of special research associate and by the internal funds of the KITS.

\end{document}